\documentclass[10pt, osa, twocolumn, nofootinbib]{revtex4-1}

\usepackage{amssymb, amsmath}

\usepackage{lineno}

\usepackage{array}
  \newcolumntype{C}[1]{>{\centering\let\newline\\\arraybackslash\hspace{0pt}}m{#1}}

\usepackage{graphicx}

\usepackage{siunitx}
    \DeclareSIUnit{\molar}{M}
    \DeclareSIUnit{\pixel}{px}
    
\usepackage[pdfborder={0 0 0}, pdftex]{hyperref}

\begin{document}


\title{The Effective Interaction Strength in a \\ Bose-Einstein Condensate of Photons in a Dye-Filled Microcavity}

\author{S. Greveling}
\affiliation{Debye Institute for Nanomaterials Science $\&$ Center for Extreme Matter and Emergent Phenomena, Utrecht University, Princetonplein 5, 3584 CC Utrecht, The Netherlands}

\author{F. van der Laan}
\affiliation{Debye Institute for Nanomaterials Science $\&$ Center for Extreme Matter and Emergent Phenomena, Utrecht University, Princetonplein 5, 3584 CC Utrecht, The Netherlands}

\author{K. L. Perrier}
\affiliation{Debye Institute for Nanomaterials Science $\&$ Center for Extreme Matter and Emergent Phenomena, Utrecht University, Princetonplein 5, 3584 CC Utrecht, The Netherlands}

\author{D. van Oosten}
\email[Corresponding author: ]{D.vanOosten@uu.nl}
\affiliation{Debye Institute for Nanomaterials Science $\&$ Center for Extreme Matter and Emergent Phenomena, Utrecht University, Princetonplein 5, 3584 CC Utrecht, The Netherlands}

\begin{abstract}
We experimentally study Bose-Einstein condensation of photons (phBEC) in a dye-filled microcavity. Through multiple absorption and emission cycles the photons inside the microcavity thermalize to the rovibronic temperature of the dye solution. Raising the photon density of the thermalized photon gas beyond the critical photon density yields a macroscopic occupation of the ground state,~\textit{i.e.} phBEC. For increasing density, we observe an increase of the condensate radius which we attribute to effective repulsive interactions. For several dye concentrations we accurately determine the radius of the condensate as a function of the number of condensate photons, and derive an effective interaction strength parameter $\tilde{g}$. For all concentrations we find $\tilde{g} \sim \num{10E-2}$, one order larger than previously reported.
\end{abstract}

\date{\today}
\maketitle

\textit{Introduction \label{sec:introduction}} --- Bose-Einstein condensates (BEC) have been formed using cold atomic gases, magnons, and exciton-polaritons as bosonic constituents~\cite{Anderson1995, Ketterle1995, Demokritov2006, Kasprzak2006, Snoke2007}. More recently, phBEC has also been achieved using a dye-filled microcavity~\cite{Klaers2010, Nyman2016}. BEC is closely related to the phenomenon of superfluidity. However, whereas the condensate is related to equilibrium properties, the superfluid is related to transport properties. Moreover, while BEC can be achieved in an ideal gas, interparticle interactions are required for superfluidity. A crucial question is therefore, in a phBEC do photon-photon interactions exist, and if so, are they attractive or repulsive?

In the case for exciton-polariton condensates it has recently been claimed by Sun~\textit{et al.}~\cite{Sun2017} that polaritons are in the strongly interacting regime with an interaction strength two orders of magnitude larger than previous experimental determinations and theoretical estimates. Previous work on the subject of interactions by Klaers~\textit{et al.}~\cite{Klaers2010} and Marelic~\textit{et al.}~\cite{Nyman2016} suggests there are effective repulsive interactions for a phBEC. Klaers~\textit{et al.}~\cite{Klaers2010} estimate an effective dimensionless interaction strength of $\tilde{g} \approx \num{7.5E-4}$ from the increase in the condensate radius as function of the number of condensate photons. Marelic~\textit{et al.}~\cite{Nyman2016} measured the quasi-particle dispersion in the condensate, from which they determine an upper limit of $\tilde{g} \leq \num{E-3}$. Several interaction mechanisms are discussed in the literature~\cite{Klaers2011, Nyman2014, Wurff2014}, where different models yield $10^{-4} < \tilde{g} < 10^{-9}$. A more accurate determination of $\tilde{g}$ is essential.

In this letter, we carefully measure the condensate radius and condensate photon number for a large number of pump powers and several dye concentrations. Images of the condensate are taken on a single-shot basis, taking special care to alternate shots of high and low pump powers to minimize possible cumulative thermal effects. We develop a theoretical model to accurately fit the condensate radius as well as the thermal tail of the photon density distribution, which is essential for an accurate determination of the condensate number. Finally, we determine $\tilde{g}$ for each dye concentration. We find values more than one order of magnitude larger than previously reported.

\bigbreak

\textit{Setup   \label{sec:setup}} --- In our setup light is confined in a microcavity consisting of two curved mirrors. Each mirror has a reflectivity of $\SI{99.9985}{\percent}$, and a radius of curvature of \SI{1}{\meter}. The typical cavity length is \SI{2}{\micro \meter}. Between the mirrors a droplet of Rhodamine 6G dissolved in ethylene glycol is held in place by the capillary force. 
\begin{figure}[!b]
  \centering
  \includegraphics[width=0.95\linewidth]{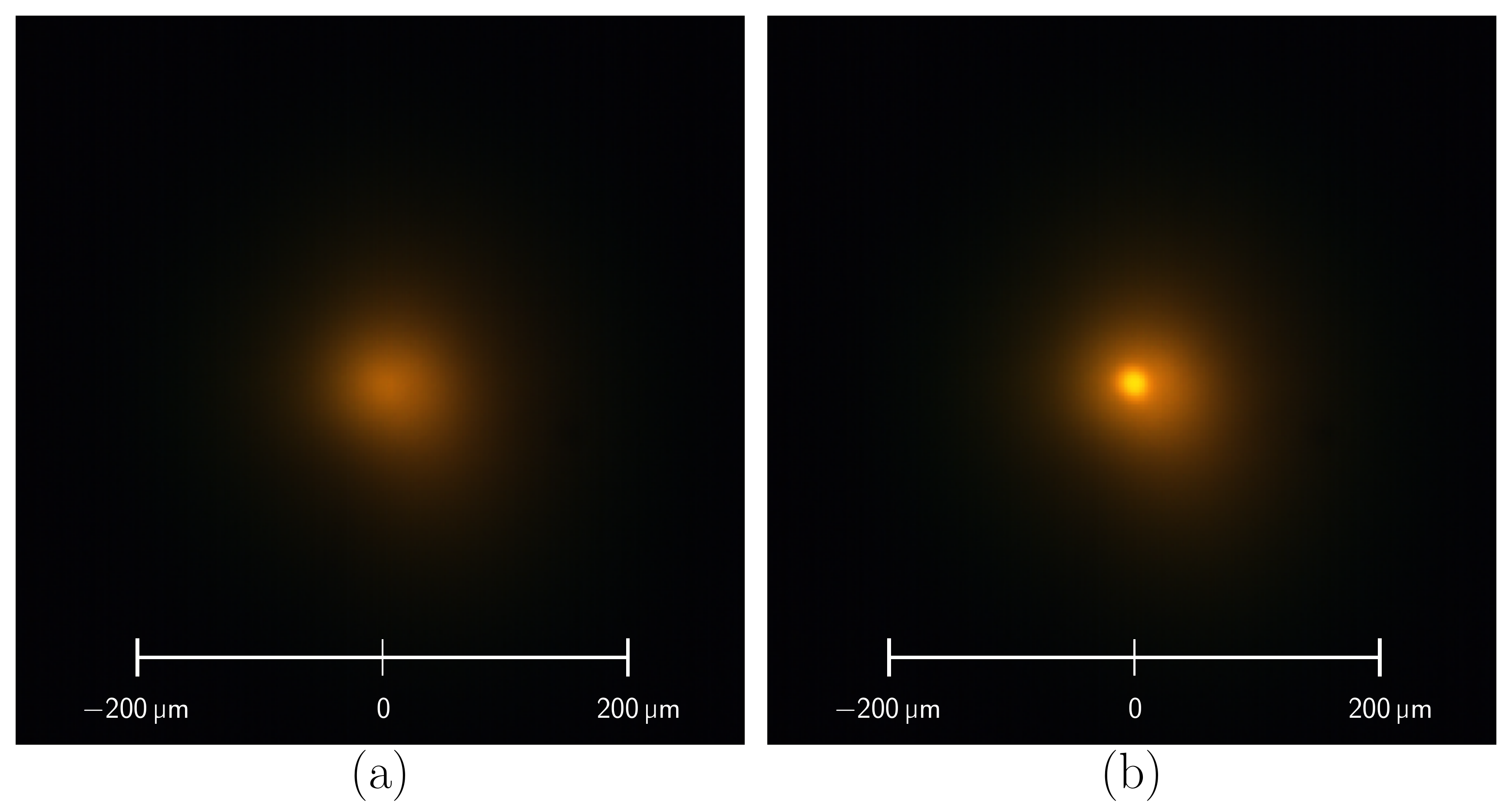}
  \caption{Spatial distribution of the photons emitted along the cavity axis as imaged on a color CCD camera. (a) Below the critical photon number, a non-condensed photon gas is obtained. (b) Above the critical photon number a phBEC as indicated by the bright yellow signal, is created at the center of the trap.    \label{fig:Figure_1}}
\end{figure}
The dye molecules are excited using a \SI{500}{\nano \second} laser pulse  with a wavelength of \SI{532}{\nano \meter} directed under an angle of \SI{65}{\degree} to the optical axis of the cavity, exploiting a reflectivity minimum of the mirrors.

Light inside the microcavity interacts with the dye molecules. with a wavelength of \SI{532}{\nano \meter} directed under an angle of \SI{65}{\degree} to the optical axis of the cavity, exploiting a reflectivity minimum of the mirrors. The setup is based on that of Klaers~\textit{et al.}~\cite{Klaers2010, Klaers2011}. Photons inside the microcavity interact with the dye molecules. Through repeated absorption and emission cycles, the trapped photons thermalize to the rovibronic temperature of the dye solution~\cite{Klaers2010, Keeling2013}. The resulting spatial distribution is shown in Fig.~\ref{fig:Figure_1}(a). Here, the photons within the signal periphery are blue-shifted due to the curvature of the mirrors. When the photon number is increased beyond the critical photon number, the additional photons occupy the ground state of the system forming a phBEC at the center of the trap, which we identify by the bright yellow ``cherry-pit" in the center of Fig.~\ref{fig:Figure_1}(b).

\bigbreak

\textit{Experiment  \label{sec:experiment}} --- We determine the size of the condensate as function of the number of condensate photons. An increase or decrease in condensate radius for increasing numbers of condensate photons can be interpreted as effective repulsive or attractive interactions, respectively. During the experiment the photon density inside the cavity is imaged for each individual pump pulse with a repetition rate of \SI{8}{\Hz}, using a 5.5 megapixel sCMOS camera with a dynamic range of 16 bits\footnote{Andor, Zyla 5.5 sCMOS}. The imaging scale is \SI{485}{\nm \per \pixel}.

For each shot of the experiment, the number of photons in the cavity is controlled by the power of the pump pulse. To exclude cumulative heating effects, the sequence of shots is chosen such that high and low pump powers always alternate, ensuring that the total pump power of two consecutive shots remains constant. The experimental sequence is performed for a total of four different concentrations of Rhodamine 6G; \num{1.5}, \num{6.0}, \num{10.5}, and \SI{14.9}{\milli \molar}.

\bigbreak

\begin{figure}[!b]
  \centering
  \includegraphics[width=0.95\linewidth]{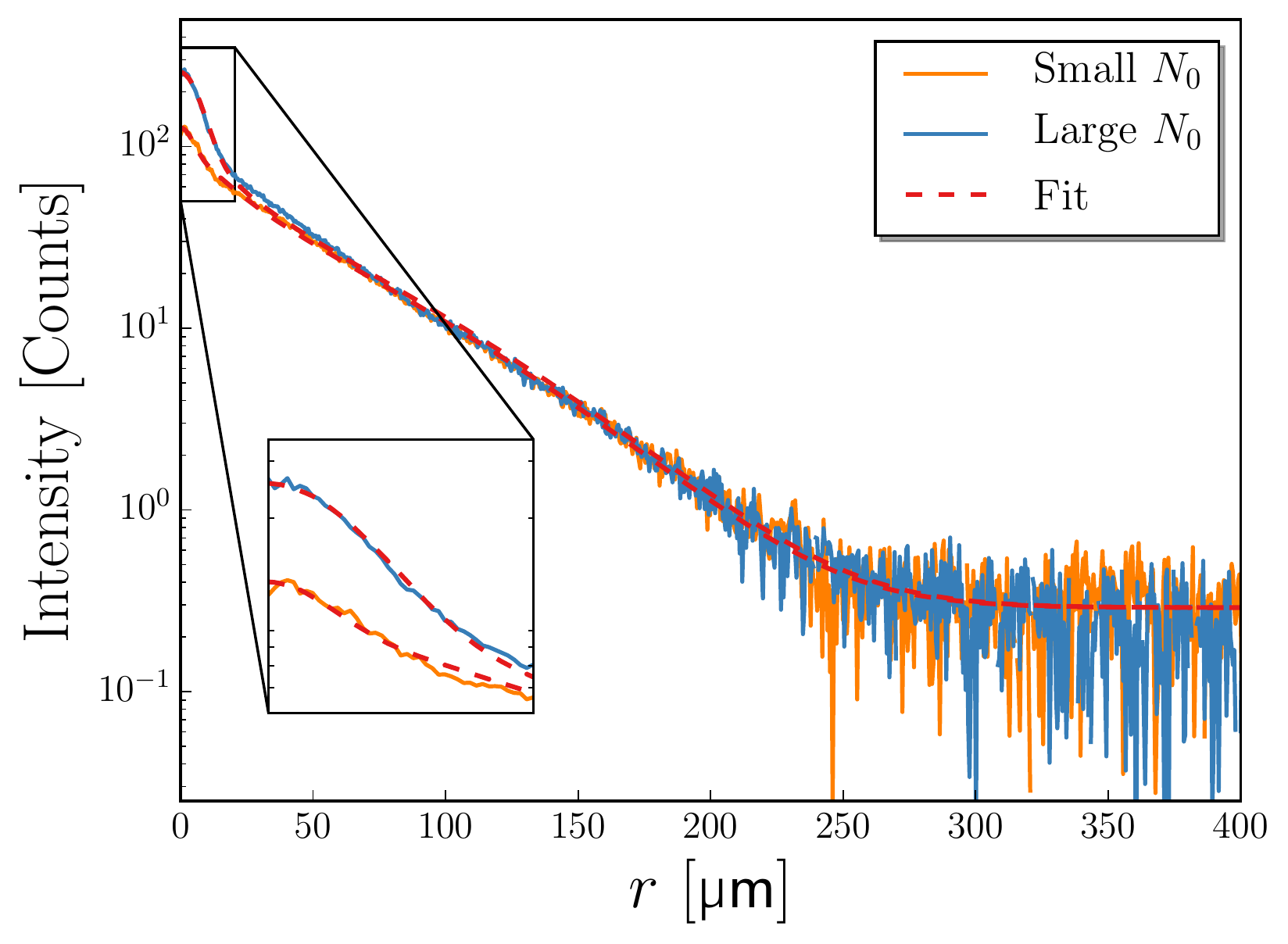}
  \caption{Radially averaged profile of the spatial distribution of the photon condensate for small (orange), and large (blue) number of photons in the condensate $N_{\mathrm{0}}$. The red dashed lines indicate a fit to the radial profiles.   \label{fig:Figure_2}}
\end{figure}

\textit{Analysis   \label{sec:analysis}} --- To determine the size of the condensate we first radially average the spatial distribution of the experimental signal. Examples of such radial averages are shown in Fig.~\ref{fig:Figure_2} for two different condensate fractions $N_{\mathrm{0}}/N_{\mathrm{tot}}$, where $N_{\mathrm{0}}$ denotes the number of photons in the condensate and $N_{\mathrm{tot}}$ the total number of photons in the system. In Fig.~\ref{fig:Figure_2}
at the trap center,~\textit{i.e.} low $r$, a decrease of the signal is observed for increasing $r$, since the size of the condensate is finite. From $r \approx \SI{20}{\micro \meter}$, the signal is exponential indicating the thermal cloud. Finally, at large $r$ a signal offset and background noise are observed. More importantly, if the radial profiles for small and large $N_{\mathrm{0}}$ are compared to each other, an increase in the radius of the condensate is observed for larger values of $N_{\mathrm{0}}$.

To determine the size of the condensate we fit the theoretical model described by Ref.~\cite{greveling2017} to the radially averaged spatial distributions keeping the harmonic oscillator length $l_{\mathrm{HO}}$ fixed. To take interaction effects in the condensate into account we allow the ground state density distribution to have a different spatial scale, $R_{\mathrm{c}} = \alpha l_{\mathrm{HO}}$ where $\alpha$ denotes a spatial scale factor:
\begin{equation}    \label{eq:rad_theory}
  \begin{split}
    \rho(r) = g_{0}\rho_{0}(\alpha r)f_{\mathrm{BE}}(E_{0},\mu,k_{\mathrm{B}}T)\eta(E_{0}) \\ 
    + \sum_{n \geq 1}^{n} g_{n} \rho_{n}(r) f_{\rm BE}(E_{n},\mu,k_{\mathrm{B}}T) \eta(E_{n}),
  \end{split}
\end{equation}
where the radial coordinate is expressed in terms of $l_{\mathrm{HO}}$ as~\cite{Klaers2011}
\begin{equation}
  l_{\mathrm{HO}} = \sqrt{ \frac{\hbar}{m_{\mathrm{ph}}\Omega} },
\end{equation}
where $m_{\mathrm{ph}} = \SI{7.7E-36}{\kg}$ denotes the effective photon mass, and $\Omega = 2\pi \cdot \SI{2.1E11}{\Hz}$ the trapping frequency.

The fit parameters in Eq.~\ref{eq:rad_theory} are the temperature $T$, the chemical potential $\mu$ and the spatial scale factor $\alpha$. The temperature determines the slope of the thermal cloud, whereas the chemical potential determines, for a given temperature the height of the condensate peak. The factor $\alpha$ scales the width of the condensate peak.

In Fig.~\ref{fig:Figure_2} the fit results to the radial profiles for both the small and large $N_{\mathrm{0}}$ are indicated by the red dashed curves. The model fits the spatial distribution of the photon condensate extremely well with a reduced $\chi^{2}$ of \num{7E-4}.

\bigbreak

\begin{figure}[!b]
  \centering
  \includegraphics[width=0.95\linewidth]{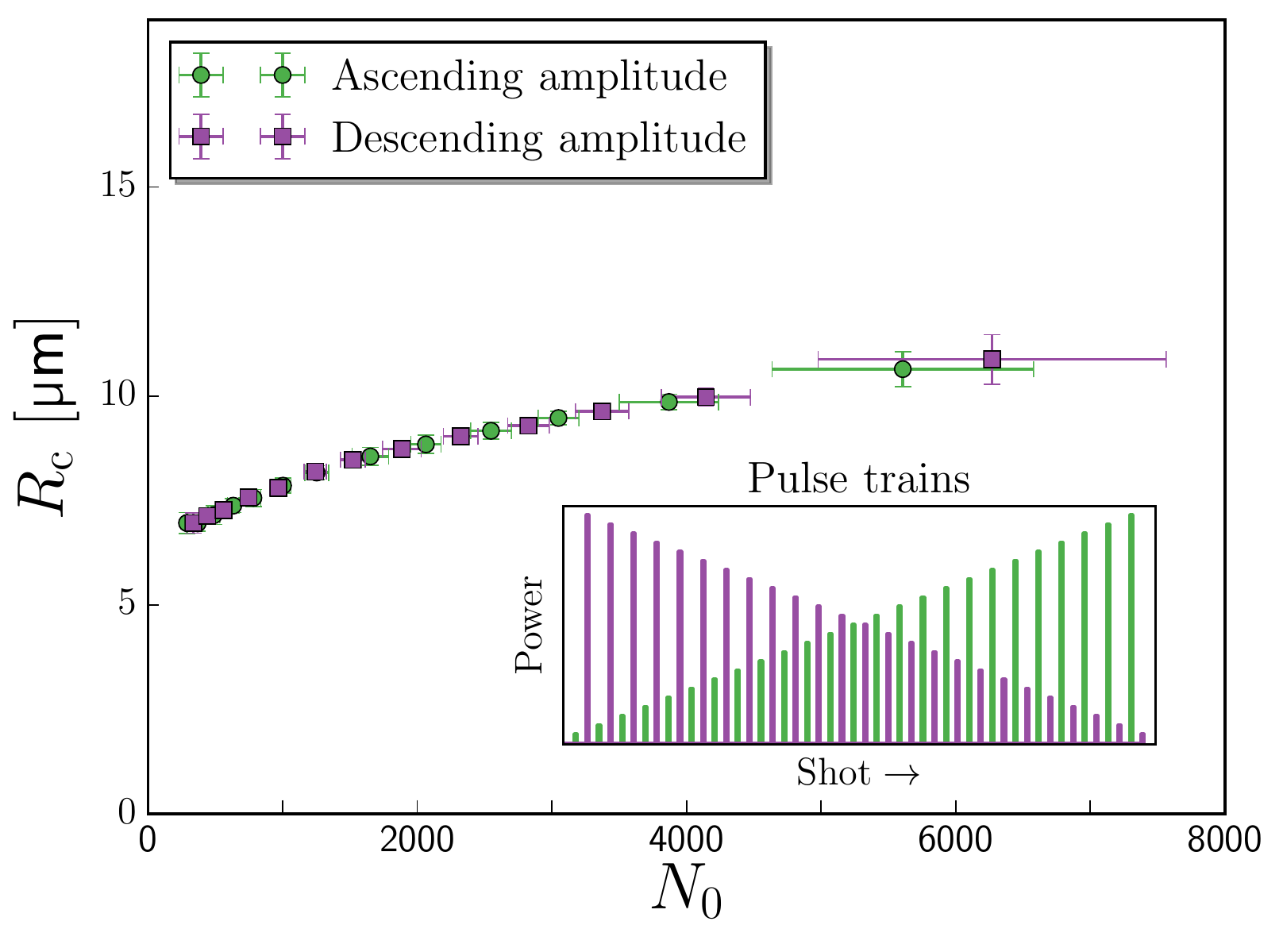}
  \caption{The condensate radius $R_{\mathrm{c}}$ as a function of the number of condensate photons $N_{0}$ for both ascending (green circles) and descending (purple squares) amplitudes. In the inset the interleaved pulse trains are shown as used in the experimental sequence. \label{fig:Figure_3}}
\end{figure}

\textit{Results  \label{sec:results}} --- The results of the analysis as described in Sec.~\ref{sec:analysis} are shown in Fig.~\ref{fig:Figure_3}, where $R_{\mathrm{c}}$ is plotted as a function of $N_{\mathrm{0}}$. 
As mentioned in Sec.~{sec:experiment}, the sequence is chosen such that high and low pulse powers alternate. As shown in the inset of Fig.~\ref{fig:Figure_3}, this effectively results in two interleaved pulse trains, one with ascending and one with descending amplitude. Each data point in Fig.~\ref{fig:Figure_3} is an average over \num{100} fit results of individual spatial distributions. For increasing $N_{\mathrm{0}}$ the size $R_{\mathrm{c}}$ increases,~\textit{i.e.}, the condensate grows which we attribute to effective repulsive interactions. 

From Fig.~\ref{fig:Figure_3} one can observe that the increase in condensate radius is identical for both the ascending and descending amplitudes. We conclude that no cumulative effects play a role and thus that the mechanism that causes the increase in condensate radius takes place within the individual shots of the experiment, hence on a time scale shorter than \SI{500}{\nano \second}. 

\begin{figure}[!b]
  \centering
  \includegraphics[width=0.95\linewidth]{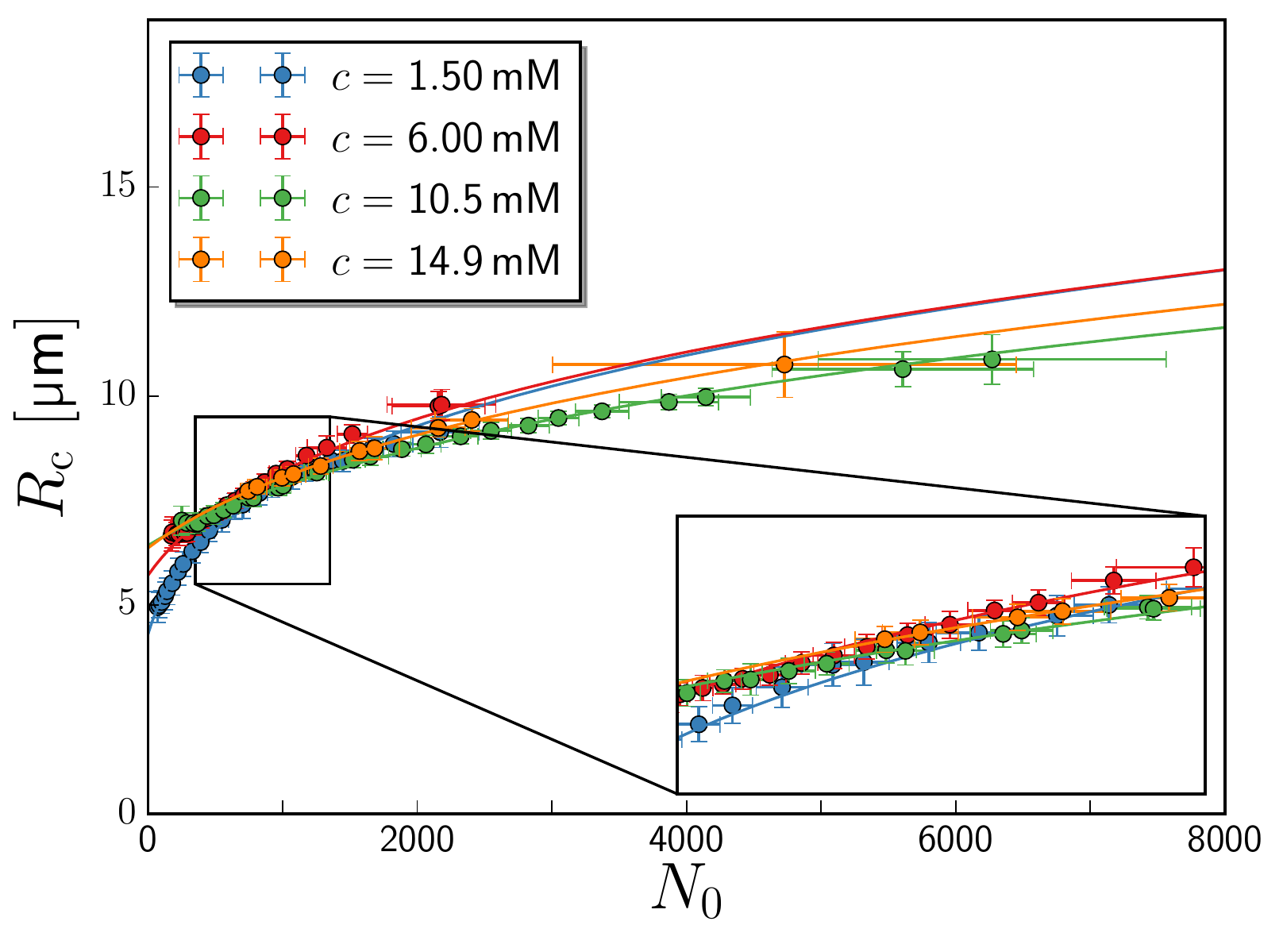}
  \caption{The condensate radius $R_{\mathrm{c}}$ as function of the number of condensate photons $N_{0}$ for four different dye concentrations. The continuous lines are fits for each dye concentration according to Eq.~\ref{eq:int_strength} to determine the dimensionless interaction strength $\tilde{g}$.   \label{fig:Figure_4}}
\end{figure}

When performing the experimental sequence for different dye concentrations we observe that the growth behavior of the condensate is similar. For every dye concentration the results of the ascending and descending amplitudes are the same. Therefore, we combine the results of the train of pulses into one data set as shown in Fig.~\ref{fig:Figure_4}. Here, $R_{\mathrm{c}}$ is plotted as function of $N_{\mathrm{0}}$ for the four different dye concentrations.
An increase in the radius of the condensate is observed for increasing numbers of condensate photons for every dye concentration. We also observe that the dependence of the condensate radius on the condensate photon number depends on the dye concentration. 

Although the origin of the interactions is unknown, a possible mechanism is described in Ref.~\cite{Wurff2014}. The grand potential of the photon gas is determined by a variational approach and by assuming the effective (long-wavelength) interactions to be a contact interaction. Wurff~\textit{et al.}~\cite{Wurff2014} find that the condensate radius is given by  
\begin{equation}    \label{eq:int_strength}
  R_{\mathrm{c}} = l_{\mathrm{HO}} \sqrt[\uproot{2}\scriptstyle 4]{1 + \frac{\tilde{g}N_{\mathrm{0}}}{2 \pi}},
\end{equation}
where $\tilde{g} = mg/\hbar^{2}$ denotes a dimensionless interaction strength. Here, $m$ is the mass of the particles,~\textit{i.e.} the effective mass of the photons in the cavity, and $g$ the coupling constant of the effective two-dimensional pointlike interaction between the particles. Using this relation, we can obtain a measure for the effective interaction strength for each dye concentration.

The colored lines in Fig.~\ref{fig:Figure_4} show the fit results of Eq.~\ref{eq:int_strength} to the data for each dye concentration. to the data for each dye concentration. The results for $\tilde{g}$ and their corresponding standard deviation are listed in Table~\ref{tab:table_1}.
\begin{center}
\begin{table}[!b]
  \begin{tabular}{| C{0.25\linewidth} | C{0.45\linewidth} |}
      \hline
    $c\,\,[\si{\milli \molar}]$     & $\tilde{g}$         \\
      \hline 
    $1.50$                          & $\num{12.4(13)E-2}$ \\
    $6.00$                          & $\num{4.1(3)E-2}$   \\
    $10.5$                          & $\num{1.5(1)E-2}$   \\
    $14.9$                          & $\num{2.0(1)E-2}$   \\
      \hline
  \end{tabular}
  \caption{Dimensionless effective interaction strength $\tilde{g}$ for different dye concentrations. \label{tab:table_1}}
\end{table}
\end{center}
Here we find a carefully determined effective interaction strength for each dye concentration. The obtained effective interaction strengths are more than one order larger than previously experimentally determined and theoretically predicted. We observe that for decreasing dye concentrations, the interaction strength increases. A possible mechanism is described in the supplementary material of Ref.~\cite{Wurff2014}, in which it is shown that the interaction strength will decrease with increasing dye concentration, in accordance with our experimental observation.

\bigbreak

\textit{Conclusion		\label{sec:conclusion}} --- We show that for a phBEC in a dye-filled microcavity the condensate radius increases for increasing numbers of condensate photons, indicating effective repulsive interactions between the photons. By interleaving positive and negative power ramps in the experimental sequence, we can exclude cumulative effects as a cause of the radius increase and thus put an upper limit of \SI{500}{\nano \second} on the timescale of the effective interactions.

We find an effective interaction strength more than one order of magnitude larger than previously experimentally determined~\cite{Klaers2010}. The effective interaction strength decreases with increasing dye concentration, in accordance with theoretical predictions. Besides the previously experimental determinations, the effective interaction strengths are also larger than theoretically expected which can be due to a simplified two-level model of the dye molecules used in the theoretical models.

The discrepancy with our findings and the upper limit found in the dynamic experiments of Marelic~\textit{et al.}~\cite{Nyman2016}, shows that care needs to be taken in regards to the time scale of the effective interactions. This will be the topic of future research. Understanding the time scale of the effective interactions will lead to insight in the mechanism behind them. If the effective interactions are of a photon-photon nature, an interaction strength this large suggests that the observation of superfluidity or even BKT physics is within reach. 

\bigbreak

\textit{Acknowledgements} --- It is a pleasure to thank Javier Hernandez Rueda, Erik van der Wurff, and Henk Stoof for usefull discussions. This work is part of the Netherlands Organization for Scientific Research (NWO).

\end{document}